\documentstyle[12pt]{article}
\newlength{\pubnumber} 

\catcode`\@=11
\@addtoreset{equation}{section}

\def\section{\@startsection{section}{1}{\z@}{3.5ex plus 1ex minus .2ex}
 {2.3ex plus .2ex}{\large\bf}}
\def\subsection{\@startsection{subsection}{2}{\z@}{2.3ex plus .2ex}
 {2.3ex plus .2ex}{\bf}}

\begin{document}

\begin{titlepage}
\samepage{
\setcounter{page}{1}
\rightline{OUTP--00--34P}
\rightline{\tt hep-ph/0107094}
\rightline{July 2001}
\vfill
\begin{center}
 {\Large \bf Doublet--Triplet Splitting\\ in Realistic Heterotic String
Derived models\\}
\vfill
\vspace{.25in}
 {\large Alon E. Faraggi\footnote{faraggi@thphys.ox.ac.uk}\\}
\vspace{.25in}
 {\it  Theoretical Physics Department,\\
              University of Oxford, Oxford, OX1 3NP, United Kingdom\\}
\vspace{.25in} {\it and}\\
\vspace{.25in}
{\it Theory Division, CERN, CH--1211 Geneva, Switzerland}
\vspace{.05in}
\end{center}
\vfill
\begin{abstract}
  {\rm
There has been recently a surge of interest in Grand Unified Theories
on orbifolds of higher dimensional spaces. In particular, the higher
dimensional doublet--triplet splitting mechanism has been of much interest.
I revisit the superstring doublet--triplet splitting mechanism
in which the color triplets are projected out by the GSO projections, 
while leaving the electroweak doublets in the physical spectrum.
The connection with the higher dimensional theories is elucidated.
It is shown that the doublet--triplet splitting depends crucially
on the assignment of boundary conditions in the compactified directions.
The possibility of reducing the number of Higgs multiplets by using the
GSO projections is also discussed.
}
\end{abstract}
\vfill
\smallskip}
\end{titlepage}

\setcounter{footnote}{0}

\def\beq{\begin{equation}}
\def\eeq{\end{equation}}
\def\beqn{\begin{eqnarray}}
\def\eeqn{\end{eqnarray}}
\def\Tr{{\rm Tr}\,}
\def\KM{{Ka\v{c}-Moody}}

\def\ie{{\it i.e.}}
\def\etc{{\it etc}}
\def\eg{{\it e.g.}}
\def\half{{\textstyle{1\over 2}}}
\def\third{{\textstyle {1\over3}}}
\def\quarter{{\textstyle {1\over4}}}
\def\m{{\tt -}}
\def\p{{\tt +}}

\def\rep#1{{\bf {#1}}}
\def\slash#1{#1\hskip-6pt/\hskip6pt}
\def\slk{\slash{k}}
\def\GeV{\,{\rm GeV}}
\def\TeV{\,{\rm TeV}}
\def\y{\,{\rm y}}
\def\SM{Standard-Model }
\def\SUSY{supersymmetry }
\def\SSM{supersymmetric standard model}
\def\vev#1{\left\langle #1\right\rangle}
\def\l{\langle}
\def\r{\rangle}

\def\Htw{{\tilde H}}
\def\chibar{{\overline{\chi}}}
\def\qbar{{\overline{q}}}
\def\ibar{{\overline{\imath}}}
\def\jbar{{\overline{\jmath}}}
\def\Hbar{{\overline{H}}}
\def\Qbar{{\overline{Q}}}
\def\abar{{\overline{a}}}
\def\alphabar{{\overline{\alpha}}}
\def\betabar{{\overline{\beta}}}
\def\tautwo{{ \tau_2 }}
\def\calF{{\cal F}}
\def\calP{{\cal P}}
\def\calN{{\cal N}}
\def\smallmatrix#1#2#3#4{{ {{#1}~{#2}\choose{#3}~{#4}} }}
\def\bone{{\bf 1}}
\def\V{{\bf V}}
\def\b{{\bf b}}
\def\N{{\bf N}}
\def\bQ{{\bf Q}}
\def\t#1#2{{ \Theta\left\lbrack \matrix{ {#1}\cr {#2}\cr }\right\rbrack }}
\def\C#1#2{{ C\left\lbrack \matrix{ {#1}\cr {#2}\cr }\right\rbrack }}
\def\tp#1#2{{ \Theta'\left\lbrack \matrix{ {#1}\cr {#2}\cr }\right\rbrack }}
\def\tpp#1#2{{ \Theta''\left\lbrack \matrix{ {#1}\cr {#2}\cr }\right\rbrack }}


\def\inbar{\,\vrule height1.5ex width.4pt depth0pt}

\def\IC{\relax\hbox{$\inbar\kern-.3em{\rm C}$}}
\def\IQ{\relax\hbox{$\inbar\kern-.3em{\rm Q}$}}
\def\IR{\relax{\rm I\kern-.18em R}}
 \font\cmss=cmss10 \font\cmsss=cmss10 at 7pt
\def\IZ{\relax\ifmmode\mathchoice
 {\hbox{\cmss Z\kern-.4em Z}}{\hbox{\cmss Z\kern-.4em Z}}
 {\lower.9pt\hbox{\cmsss Z\kern-.4em Z}}
 {\lower1.2pt\hbox{\cmsss Z\kern-.4em Z}}\else{\cmss Z\kern-.4em Z}\fi}

\def\AEF{A.E. Faraggi}
\def\KRD{K.R. Dienes}
\def\JMR{J. March-Russell}
\def\NPB#1#2#3{{\it Nucl.\ Phys.}\/ {\bf B#1} (#2) #3}
\def\PLB#1#2#3{{\it Phys.\ Lett.}\/ {\bf B#1} (#2) #3}
\def\PRD#1#2#3{{\it Phys.\ Rev.}\/ {\bf D#1} (#2) #3}
\def\PRL#1#2#3{{\it Phys.\ Rev.\ Lett.}\/ {\bf #1} (#2) #3}
\def\PRT#1#2#3{{\it Phys.\ Rep.}\/ {\bf#1} (#2) #3}
\def\MODA#1#2#3{{\it Mod.\ Phys.\ Lett.}\/ {\bf A#1} (#2) #3}
\def\IJMP#1#2#3{{\it Int.\ J.\ Mod.\ Phys.}\/ {\bf A#1} (#2) #3}
\def\nuvc#1#2#3{{\it Nuovo Cimento}\/ {\bf #1A} (#2) #3}
\def\etal{{\it et al,\/}\ }

\hyphenation{su-per-sym-met-ric non-su-per-sym-met-ric}
\hyphenation{space-time-super-sym-met-ric}
\hyphenation{mod-u-lar mod-u-lar--in-var-i-ant}


\setcounter{footnote}{0}
\section{Introduction}\label{intro}

Superstring theory provides a consistent framework for perturbative
unification of gravity and gauge theories. Among the five perturbative string
limits the heterotic string is the only one that also admits the
standard grand unification structures, for example that of $SO(10)$.
The specific implications of this is that the heterotic string can
in principle preserve the embedding of the standard model generations
in the 16 representation of $SO(10)$ as well as the canonical normalization
of the weak hypercharge. However, as is well known supersymmetric 
grand unified theories \cite{nilles} give rise to proton decay from
dimension four, five and six operators \cite{WSY}.

Superstring theory offers resolutions to the proton decay from all
of these sources. The issue of proton decay in realistic string
constructions has been amply discussed in the past
\cite{psinsm,ps94,cus,pati,zprime} and therefore 
I will be brief in respect to the details that are given in the
earlier literature. The purpose of this note, prompted by the
recent interest in grand unified higher dimensional theories \cite{toons},
is to elucidate the connection between the superstring doublet--triplet
splitting mechanism, which was derived in the free fermionic formulation 
and the higher dimensional theories.

\setcounter{footnote}{0}
\section{Realistic free fermionic models}\label{construct}

The class of models under consideration are constructed in the
free fermionic formulation \cite{fff}. The notation 
and details of the construction of these
models are given elsewhere \cite{fsu5,fny,alr,eu,slm,nahe,cfn,cfs}.
In the free fermionic formulation of the heterotic string
in four dimensions all the world--sheet
degrees of freedom  required to cancel
the conformal anomaly are represented in terms of free fermions
propagating on the string world--sheet.
In the light--cone gauge the world--sheet field content consists
of two transverse left-- and right--moving space--time coordinate bosons,
$X_{1,2}^\mu$ and ${\bar X}_{1,2}^\mu$,
and their left--moving fermionic superpartners $\psi^\mu_{1,2}$,
and additional 62 purely internal
Majorana--Weyl fermions, of which 18 are left--moving,
$\chi^{I}$, and 44 are right--moving, $\phi^a$.
In the supersymmetric sector the world--sheet supersymmetry is realized
non--linearly and the world--sheet supercurrent is given by
$T_F=\psi^\mu\partial X_\mu+i\chi^Iy^I\omega^I,~(I=1,\cdots,6).$
The $\{\chi^{I},y^I,\omega^I\}~(I=1,\cdots,6)$ are 18 real free
fermions transforming as the adjoint representation of $SU(2)^6$.
Under parallel transport around a noncontractible loop on the toroidal
world--sheet the fermionic fields pick up a phase
\begin{equation}
f~\rightarrow~-{\rm e}^{i\pi\alpha(f)}f~,~~\alpha(f)\in(-1,+1].
\label{fermionphase}
\end{equation}
Each set of specified
phases for all world--sheet fermions, around all the non--contractible
loops is called the spin structure of the model. Such spin structures
are usually given is the form of 64 dimensional boundary condition vectors,
with each element of the vector specifying the phase of the corresponding
world--sheet fermion. The basis vectors are constrained by string consistency
requirements and completely determine the vacuum structure of the model.
The physical spectrum is obtained by applying the generalized GSO projections.
The low energy effective field theory is obtained by S--matrix elements
between external states \cite{kln}.  

The boundary condition basis defining a typical 
realistic free fermionic heterotic string models is 
constructed in two stages. 
The first stage consists of the NAHE set,
which is a set of five boundary condition basis vectors, 
$\{{\bf1},S,b_1,b_2,b_3\}$ \cite{nahe}. 
The gauge group after imposing the GSO projections induced
by the NAHE set is $SO(10)\times SO(6)^3\times E_8$
with $N=1$ supersymmetry. The space--time vector bosons that generate
the gauge group arise from the Neveu--Schwarz sector and
from the sector ${\bf1}+b_1+b_2+b_3$. The Neveu--Schwarz sector
produces the generators of $SO(10)\times SO(6)^3\times SO(16)$.
The sector $\zeta\equiv{\bf1}+b_1+b_2+b_3$ produces the spinorial  
${\bf128}$
of $SO(16)$ and completes the hidden gauge group to $E_8$.
The NAHE set divides the internal world--sheet 
fermions in the following way: ${\bar\phi}^{1,\cdots,8}$ generate the 
hidden $E_8$ gauge group, ${\bar\psi}^{1,\cdots,5}$ generate the $SO(10)$ 
gauge group, and $\{{\bar y}^{3,\cdots,6},{\bar\eta}^1\}$,  
$\{{\bar y}^1,{\bar y}^2,{\bar\omega}^5,{\bar\omega}^6,{\bar\eta}^2\}$,
$\{{\bar\omega}^{1,\cdots,4},{\bar\eta}^3\}$ generate the three horizontal 
$SO(6)^3$ symmetries. The left--moving $\{y,\omega\}$ states are divided 
to $\{{y}^{3,\cdots,6}\}$,  
$\{{y}^1,{y}^2,{\omega}^5,{\omega}^6\}$,
$\{{\omega}^{1,\cdots,4}\}$ and $\chi^{12}$, $\chi^{34}$, $\chi^{56}$ 
generate the left--moving $N=2$ world--sheet supersymmetry.
At the level of the NAHE set the sectors $b_1$, $b_2$ and $b_3$
produce 48 multiplets, 16 from each, in the $16$ 
representation of $SO(10)$. The states from the sectors $b_j$
are singlets of the hidden $E_8$ gauge group and transform 
under the horizontal $SO(6)_j$ $(j=1,2,3)$ symmetries. This structure
is common to all the realistic free fermionic models.

The second stage of the
basis construction consists of adding to the 
NAHE set three (or four) additional boundary condition basis vectors. 
These additional basis vectors reduce the number of generations
to three chiral generations, one from each of the sectors $b_1$,
$b_2$ and $b_3$, and simultaneously break the four dimensional
gauge group. The assignment of boundary conditions to
$\{{\bar\psi}^{1,\cdots,5}\}$ breaks $SO(10)$ to one of its subgroups
$SU(5)\times U(1)$ \cite{fsu5}, $SO(6)\times SO(4)$ \cite{alr},
$SU(3)\times SU(2)\times U(1)^2$ \cite{fny,eu,slm,cfn}
or $SU(3)\times SU(2)^2\times U(1)$ \cite{cfs}.
Similarly, the hidden $E_8$ symmetry is broken to one of its
subgroups by the basis vectors which extend the NAHE set.
The flavor $SO(6)^3$ symmetries in the NAHE--based models
are always broken to flavor $U(1)$ symmetries, as the breaking
of these symmetries is correlated with the number of chiral
generations. Three such $U(1)_j$ symmetries are always obtained
in the NAHE based free fermionic models, from the subgroup
of the observable $E_8$, which is orthogonal to $SO(10)$.
These are produced by the world--sheet currents ${\bar\eta}{\bar\eta}^*$
($j=1,2,3$), which are part of the Cartan sub--algebra of the
observable $E_8$. Additional unbroken $U(1)$ symmetries, denoted
typically by $U(1)_j$ ($j=4,5,...$), arise by pairing two real
fermions from the sets $\{{\bar y}^{3,\cdots,6}\}$,
$\{{\bar y}^{1,2},{\bar\omega}^{5,6}\}$ and
$\{{\bar\omega}^{1,\cdots,4}\}$. The final observable gauge
group depends on the number of such pairings. 

\setcounter{footnote}{0}
\section{Superstring Higgs doublet--triplet splitting}\label{higgs}

In the free fermionic models, representations in the $5$ and ${\bar 5}$
of $SU(5)$ which yield electroweak Higgs doublets and color Higgs triplets
arise from the untwisted (Neveu--Schwarz) and twisted sectors. The color
triplets arising from these sectors are those that can mediate rapid proton
decay from dimension five operators. Additional color triplets may arise
from ``Wilsonian'' sectors, but their interactions with the Standard Model
states may be protected by discrete symmetries \cite{cus}. 

For the Higgs multiplets arising from the Neveu--Schwarz sector
there exists a doublet--triplet splitting mechanism which operates
by the assignment of boundary conditions to the set of internal 
world--sheet fermions $\{y,\omega\vert{\bar y},{\bar\omega}\}^{1,\cdots,6}$.
The Neveu--Schwarz sector gives rise to three fields in the
10 representation of $SO(10)$.  These contain the  Higgs electroweak
doublets and color triplets. Each of those is charged with respect to one
of the horizontal $U(1)$ symmetries $U(1)_{1,2,3}$.  Each one of these
multiplets is associated, by the horizontal symmetries, with one of the
twisted sectors, $b_1$, $b_2$ and $b_3$. The doublet--triplet
splitting results from the boundary condition basis vectors which breaks
the $SO(10)$ symmetry to $SO(6)\times SO(4)$. We can define a quantity
$\Delta_i$ in these basis vectors which measures the difference between the
boundary conditions assigned to the internal fermions from the set
$\{y,w\vert{\bar y},{\bar\omega}\}$ and which are periodic in the vector
$b_i$,
\begin{equation}
\Delta_i=\vert\alpha_L({\rm internal})-
\alpha_R({\rm internal})\vert=0,1~~(i=1,2,3)
\label{dts}
\end{equation}
If $\Delta_i=0$ then the Higgs triplets, $D_i$ and ${\bar D}_i$,
remain in the massless spectrum while the Higgs doublets, $h_i$ and ${\bar
h}_i$ are projected out
and the opposite occurs for $\Delta_i=1$.

Thus, the rule in Eq. (\ref{dts})
is a generic rule that can be used in the construction
of the free fermionic models. The model
of table (\ref{colorhiggs}) illustrates this rule.
In this model $\Delta_1=\Delta_2=0$ while $\Delta_3=1$. Therefore,
this model produces two pairs of color triplets and one pair of
Higgs doublets from the Neveu--Schwarz sector, $D_1$, $\bar D_1$
$D_2$, $\bar D_2$ and $h_3$, $\bar h_3$.
\beqn
 &\begin{tabular}{c|c|ccc|c|ccc|c}
 ~ & $\psi^\mu$ & $\chi^{12}$ & $\chi^{34}$ & $\chi^{56}$ &
        $\bar{\psi}^{1,...,5} $ &
        $\bar{\eta}^1 $&
        $\bar{\eta}^2 $&
        $\bar{\eta}^3 $&
        $\bar{\phi}^{1,...,8} $ \\
\hline
\hline
  ${\alpha}$  &  1 & 1&0&0 & 1~1~1~0~0 & 1 & 0 & 1 & 1~1~1~1~0~0~0~0 \\
  ${\beta}$   &  1 & 0&1&0 & 1~1~1~0~0 & 0 & 1 & 1 & 1~1~1~1~0~0~0~0 \\
  ${\gamma}$  &  1 & 0&0&1 &
                ${1\over2}$~${1\over2}$~${1\over2}$~${1\over2}$~${1\over2}$
              & ${1\over2}$ & ${1\over2}$ & ${1\over2}$ &
                ${1\over2}$~0~1~1~${1\over2}$~${1\over2}$~${1\over2}$~0 \\
\end{tabular}
   \nonumber\\
   ~  &  ~ \nonumber\\
   ~  &  ~ \nonumber\\
     &\begin{tabular}{c|c|c|c}
 ~&   $y^3{\bar y}^3$
      $y^4{\bar y}^4$
      $y^5{\bar y}^5$
      $y^6{\bar y}^6$
  &   $y^1{\bar y}^1$
      $y^2{\bar y}^2$
      $\omega^5{\bar\omega}^5$
      $\omega^6{\bar\omega}^6$
  &   $\omega^2{\omega}^3$
      $\omega^1{\bar\omega}^1$
      $\omega^4{\bar\omega}^4$
      ${\bar\omega}^2{\bar\omega}^3$ \\
\hline
\hline
$\alpha$ & 1 ~~~ 0 ~~~ 0 ~~~ 1  & 0 ~~~ 0 ~~~ 1 ~~~ 0  & 0 ~~~ 0 ~~~ 1 ~~~ 1 \\
$\beta$  & 0 ~~~ 0 ~~~ 0 ~~~ 1  & 0 ~~~ 1 ~~~ 1 ~~~ 0  & 0 ~~~ 1 ~~~ 0 ~~~ 1 \\
$\gamma$ & 1 ~~~ 1 ~~~ 0 ~~~ 0  & 1 ~~~ 1 ~~~ 0 ~~~ 0  & 0 ~~~ 0 ~~~ 0 ~~~ 1 \\
\end{tabular}
\label{colorhiggs}
\eeqn
With the choice of generalized GSO coefficients:
\beqn
&& c\left(\matrix{b_1,b_3,\alpha,\beta,\gamma\cr
                                    \alpha\cr}\right)=
-c\left(\matrix{b_2\cr
                                    \alpha\cr}\right)=
c\left(\matrix{{\bf1},b_j,\gamma\cr
                                    \beta\cr}\right)=\nonumber\\
&& c\left(\matrix{\gamma\cr
                                   b_3\cr}\right)=
-c\left(\matrix{\gamma\cr
                                    {\bf1},b_1,b_2\cr}\right)=-1
{}~~~~~~~~~~~~~~~~~~~~~(j=1,2,3), \nonumber
\eeqn
with the others specified by modular invariance and space--time
supersymmetry.
In ref. \cite{ps94} this doublet--triplet splitting mechanism
is proven in terms of the world--sheet modular invariance constraints and
the GSO projections. The constraint 
\beq
\vert\alpha_L(b_j)-\alpha_R(b_j)\vert=1~~,~~(j=1,2,3),
\label{threedoub}
\eeq
in the basis vectors that break $SO(10)$ to $SO(6)\times SO(4)$,
guarantees that the Neveu--Schwarz color triplets, $D_j$ and ${\bar D}_j$, 
are projected out and that the electroweak doublets,
$h_j$ and ${\bar h}_j$, remain in the massless spectrum. 
It is now possible to construct models in which all the color Higgs triplets
from the Neveu--Schwarz sector are projected out by the GSO projections. 
Table (\ref{table3}) provides an example of such a model. It is noted
that the NS Higgs--doublet triplet mechanism operates irrespective of the
choice of the GSO projection coefficients. 

\beqn
 &\begin{tabular}{c|c|ccc|c|ccc|c}
 ~ & $\psi^\mu$ & $\chi^{12}$ & $\chi^{34}$ & $\chi^{56}$ &
        $\bar{\psi}^{1,...,5} $ &
        $\bar{\eta}^1 $&
        $\bar{\eta}^2 $&
        $\bar{\eta}^3 $&
        $\bar{\phi}^{1,...,8} $ \\
\hline
\hline
  ${\alpha}$  &  0 & 0&0&0 & 1~1~1~0~0 & 0 & 0 & 0 & 1~1~1~1~0~0~0~0 \\
  ${\beta}$   &  0 & 0&0&0 & 1~1~1~0~0 & 0 & 0 & 0 & 1~1~1~1~0~0~0~0 \\
  ${\gamma}$  &  0 & 0&0&0 &
                ${1\over2}$~${1\over2}$~${1\over2}$~${1\over2}$~${1\over2}$
              & ${1\over2}$ & ${1\over2}$ & ${1\over2}$ &
                ${1\over2}$~0~1~1~${1\over2}$~${1\over2}$~${1\over2}$~0 \\
\end{tabular}
   \nonumber\\
   ~  &  ~ \nonumber\\
   ~  &  ~ \nonumber\\
     &\begin{tabular}{c|c|c|c}
 ~&   $y^3{y}^6$
      $y^4{\bar y}^4$
      $y^5{\bar y}^5$
      ${\bar y}^3{\bar y}^6$
  &   $y^1{\omega}^5$
      $y^2{\bar y}^2$
      $\omega^6{\bar\omega}^6$
      ${\bar y}^1{\bar\omega}^5$
  &   $\omega^2{\omega}^4$
      $\omega^1{\bar\omega}^1$
      $\omega^3{\bar\omega}^3$
      ${\bar\omega}^2{\bar\omega}^4$ \\
\hline
\hline
$\alpha$ & 1 ~~~ 1 ~~~ 1 ~~~ 0  & 1 ~~~ 1 ~~~ 1 ~~~ 0  & 1 ~~~ 1 ~~~ 1 ~~~ 0 \\
$\beta$  & 0 ~~~ 1 ~~~ 0 ~~~ 1  & 0 ~~~ 1 ~~~ 0 ~~~ 1  & 1 ~~~ 0 ~~~ 0 ~~~ 0 \\
$\gamma$ & 0 ~~~ 0 ~~~ 1 ~~~ 1  & 1 ~~~ 0 ~~~ 0 ~~~ 0  & 0 ~~~ 1 ~~~ 0 ~~~ 1 \\
\end{tabular}
\label{table3}
\eeqn
As the Higgs--doublet triplet mechanism operates irrespective of
the GSO phases they are not displayed here explicitly.
The sector $b_1+b_2+\alpha+\beta$
produces additional states that transform solely under the observable
sector. In particular it can give rise to additional electroweak 
doublets and color triplets. 
The color triplets from this sector may cause problems with proton 
lifetime constraints. 
However, a similar 
doublet--triplet splitting mechanism works for this sector as well. 
There exist choices of boundary conditions for the set of
left--right symmetric internal fermions, 
$\{y,\omega\vert{\bar y},{\bar\omega}\}^{1,\cdots,6}$, for which 
the triplets are projected out and the doublets remain in the massless 
spectrum. For example, in the model of ref. \cite{eu}
this sector produces
one pair of electroweak doublets and one pair of color triplets, 
\beq
{h_{45}\equiv{[(1,0);(2,-1)]}_
{-{1\over2},-{1\over2},0,0,0,0} {\hskip .3cm}
D_{45}\equiv{[(3,-1);(1,0)]}_
{-{1\over2},-{1\over2},0,0,0,0}}
\label{m278}
\eeq
while in the model of table (\ref{table3})
this sector produces two 
pairs of electroweak doublets,
\beq
{h_{45}\equiv{[(1,0);(2,-1)]}_
{{1\over2},{1\over2},0,0,0,0} {\hskip .4cm}
{h}_{45}^\prime\equiv{[(1,0);(2,-1)]}_
{-{1\over2},-{1\over2},0,0,0,0}}
\label{m274}
\eeq
and all the color triplets, from the Neveu--Schwarz sector and the sector
$b_1+b_2+\alpha+\gamma$, are projected out
from the physical spectrum by the GSO projections. 
The two models differ only by the assignment 
of boundary conditions to the set of internal fermions, 
$\{y,\omega\vert{\bar y},{\bar\omega}\}^{1,\cdots,6}$. 
The simplicity and elegance of the superstring doublet--triplet
splitting mechanism is striking. There is no need for exotic 
representations of high dimensionality as in minimal $SU(5)$ extension
of the Standard Model \cite{dtssu5}. Moreover, the superstring 
doublet--triplet splitting mechanism does not depend on additional 
assumptions on Yukawa couplings as is required in all GUT 
doublet--triplet splitting mechanism. In the superstring 
doublet--triplet splitting mechanism the dangerous color triplets 
simply do not exist in the massless spectrum. 
Furthermore, due to discrete and custodial non-Abelian symmetries
\cite{cus} there exists examples of models in which proton decay
mediating operators are not generated. 

Another relevant question with regard to the Higgs doublet--triplet
splitting mechanism is whether it is possible to construct models in which
both the Higgs color triplets and electroweak doublets from the
Neveu--Schwarz sector are projected out by the GSO projections.
This is a viable possibility as we can choose for example
$$\Delta_j^{(\alpha)}=1 ~{\rm and}~ \Delta_j^{(\beta)}=0,$$
where $\Delta^{(\alpha,\beta)}$ are the projections due
to the basis vectors $\alpha$ and $\beta$ respectively.
This is a relevant question as the number of Higgs representations,
which generically appear in the massless spectrum,
is larger than what is allowed by the low energy phenomenology.
Consider for example the model in table (\ref{bothdandt})
\beqn
 &\begin{tabular}{c|c|ccc|c|ccc|c}
 ~ & $\psi^\mu$ & $\chi^{12}$ & $\chi^{34}$ & $\chi^{56}$ &
        $\bar{\psi}^{1,...,5} $ &
        $\bar{\eta}^1 $&
        $\bar{\eta}^2 $&
        $\bar{\eta}^3 $&
        $\bar{\phi}^{1,...,8} $ \\
\hline
\hline
  ${\alpha}$  &  1 & 1&0&0 & 1~1~1~0~0 & 1 & 0 & 1 & 1~1~1~1~0~0~0~0 \\
  ${\beta}$   &  1 & 0&1&0 & 1~1~1~0~0 & 0 & 1 & 1 & 1~1~1~1~0~0~0~0 \\
  ${\gamma}$  &  1 & 0&0&1 &
                ${1\over2}$~${1\over2}$~${1\over2}$~${1\over2}$~${1\over2}$
              & ${1\over2}$ & ${1\over2}$ & ${1\over2}$ &
                ${1\over2}$~0~1~1~${1\over2}$~${1\over2}$~${1\over2}$~0 \\
\end{tabular}
   \nonumber\\
   ~  &  ~ \nonumber\\
   ~  &  ~ \nonumber\\
     &\begin{tabular}{c|c|c|c}
 ~&   $y^3{y}^6$
      $y^4{\bar y}^4$
      $y^5{\bar y}^5$
      ${\bar y}^3{\bar y}^6$
  &   $y^1{\omega}^5$
      $y^2{\bar y}^2$
      $\omega^6{\bar\omega}^6$
      ${\bar y}^1{\bar\omega}^5$
  &   $\omega^2{\omega}^4$
      $\omega^1{\bar\omega}^1$
      $\omega^3{\bar\omega}^3$
      ${\bar\omega}^2{\bar\omega}^4$ \\
\hline
\hline
$\alpha$ & 1 ~~~ 0 ~~~ 0 ~~~ 0  & 0 ~~~ 0 ~~~ 1 ~~~ 1  & 0 ~~~ 0 ~~~ 1 ~~~ 1 \\
$\beta$  & 0 ~~~ 0 ~~~ 1 ~~~ 0  & 1 ~~~ 0 ~~~ 0 ~~~ 0  & 0 ~~~ 1 ~~~ 0 ~~~ 0 \\
$\gamma$ & 0 ~~~ 1 ~~~ 0 ~~~ 0  & 0 ~~~ 1 ~~~ 0 ~~~ 1  & 1 ~~~ 0 ~~~ 0 ~~~ 0 \\
\end{tabular}
\label{bothdandt}
\eeqn
With the choice of generalized GSO coefficients:
\beqn
&& c\left(\matrix{b_j\cr
                           S,b_j,\alpha,\beta,\gamma\cr}\right)=
c\left(\matrix{\alpha\cr
                           \alpha,\beta,\gamma\cr}\right)=\nonumber\\
&&c\left(\matrix{\beta,\cr
                                    \beta,\gamma\cr}\right)=
 c\left(\matrix{\gamma\cr
                                   {\bf1}\cr}\right)=-1\nonumber
\eeqn
(j=1,2,3), with the others specified by modular invariance and space--time
supersymmetry. In this model
$\Delta_1^{(\alpha)}=\Delta_2^{(\alpha)}=\Delta_3^{(\alpha)}=1$,
and $\Delta_1^{(\beta)}=\Delta_3^{(\beta)}=0$,
Therefore, In this model irrespective of the choice of the generalized GSO
projection coefficients, both the Higgs color triplets and electroweak
doublets associated with $b_1$ and $b_3$ are projected out by the
GSO projections. However, it is found that the combination of these
projections also results in the projection of some
of the representations from the corresponding sectors
$b_1$ and $b_3$ and therefore these sectors do not produce
the full chiral 16 of $SO(10)$. Therefore, realization of
this mutual projection of both Higgs triplets and doublets
from the Neveu--Schwarz sector requires that the chiral
generations be obtained from non--NAHE set basis vectors.

\setcounter{footnote}{0}
\section{Correspondence with orbifolds}\label{orbifold}

In the previous section it was shown that the superstring doublet--triplet
splitting mechanism depend on the assignment of boundary conditions 
to the set of internal world--sheet fermions
$\{y,\omega\vert{\bar y},{\bar\omega}\}$. 
In this section I show that this set of internal world--sheet 
fermions in fact corresponds to six internal compactified dimensions.
Consequently, in the bosonic language, {\it i.e.} in the language
of the compactified dimensions, the boundary condition of the
internal fermions translate to twisting of the internal dimensions
with orbifold fixed points. 

The correspondence with the orbifold construction is illustrated
by extending the NAHE set, $\{{\bf1},S,b_1,b_2,b_3\}$, by one additional
boundary condition basis vector \cite{foc}
\beq
X=(0,\cdots,0\vert{\underbrace{1,\cdots,1}_{{\bar\psi^{1,\cdots,5}},
{\bar\eta^{1,2,3}}}},0,\cdots,0)~.
\label{vectorx}
\eeq
with a suitable choice of the GSO projection coefficients the 
model possess an $SO(4)^3\times E_6\times U(1)^2\times E_8$ gauge group
and $N=1$ space--time supersymmetry. The matter fields
include 24 generations in 27 representations of
$E_6$, eight from each of the sectors $b_1\oplus b_1+X$,
$b_2\oplus b_2+X$ and $b_3\oplus b_3+X$.
Three additional 27 and $\overline{27}$ pairs are obtained
from the Neveu--Schwarz $\oplus~X$ sector.

To construct the model in the orbifold formulation one starts
with a model compactified on a flat torus with nontrivial background
fields \cite{narain}.
The action of the six dimensional compactified dimensions is given by
\beq
S={1\over{8\pi}}
\int{d^2\sigma({G_{ij}\partial{X^i}\partial{X^j}+B_{ij}\partial{X^i}
\partial{X^j}})}
\label{action6d}
\eeq
where 
\beq
G_{ij}={1\over2}{\sum_{I=1}^D}R_ie_i^IR_je_j^I
\label{metric}
\eeq
is the metric of the six dimensional compactified space 
and $B_{ij}=-B_{ji}$ is the antisymmetric tensor field. 
The $e^i=\{e_i^I\}$ are six linear independent vectors normalized
to $(e_i)^2=2$. 
The subset of basis vectors
\beq
\{{\bf1},S,X,I={\bf1}+b_1+b_2+b_3\}
\label{neq4set}
\eeq
generates a toroidally-compactified model with $N=4$ space--time
supersymmetry and $SO(12)\times E_8\times E_8$ gauge group.
The same model is obtained in the geometric (bosonic) language
by constructing the background fields which produce
the $SO(12)$ lattice. Taking the metric
of the six-dimensional compactified manifold
to be the Cartan matrix of $SO(12)$:
\beq
g_{ij}=\left(\matrix{~2&-1& ~0& ~0& ~0& ~0\cr%
-1& ~2&-1& ~0& ~0& ~0\cr~0&-1& ~2&-1& ~0& ~0\cr~0& ~0&-1
& ~2&-1&-1\cr ~0& ~0& ~0&-1& ~2& ~0\cr ~0& ~0& ~0&-1& ~0& ~2\cr}\right)
\label{gso12}
\eeq
and the antisymmetric tensor
\beq
b_{ij}=\cases{
g_{ij}&;\ $i>j$,\cr
0&;\ $i=j$,\cr
-g_{ij}&;\ $i<j$.\cr}
\label{bso12}
\eeq
When all the radii of the six-dimensional compactified
manifold are fixed at $R_I=\sqrt2$, it is seen that the
left-- and right--moving momenta
\beq
P^I_{R,L}=[m_i-{1\over2}(B_{ij}{\pm}G_{ij})n_j]{e_i^I}^*
\label{lrmomenta}
\eeq
reproduce all the massless root vectors in the lattice of
$SO(12)$,
where in (\ref{lrmomenta}) 
the ${e_i^I}^*$ are dual to the $e_i$, and
$e_i^*\cdot e_j=\delta_{ij}$.
 
The orbifold models are obtained by moding out the six dimensional torus
by a discrete symmetry group, P \cite{DHVW}. The allowed discrete symmetry
groups are constrained by modular invariance. The Hilbert space is obtained
by acting on the vacuum with twisted and untwisted oscillators and by 
projecting on states that are invariant under the space and group twists. 
A general left--right symmetric twist is given by
$(\theta^i_j,v^i;\Theta^I_J,V^I)$ $(i=1,\cdots,6)$ $(I=1,\cdots,16)$
and $X^i(2\pi)=\theta^i_j X^j(0)+v^i$; $X^I(2\pi)=\Theta^I_J X^J(0)+V^I$. 
The massless spectrum contains mass states from the untwisted and twisted
sectors. The untwisted sector is obtained by projecting 
on states that are invariant under the space and group twists. 
The twisted string centers around the points that are left 
fixed by the space twist. In the case of ``standard embedding"
one acts on the gauge degrees of freedom in an $SU(3)\in{E_8\times E_8}$
with the same action as on the six compactified dimensions + NSR fermions. 
In this case the number of chiral families (27's of $E_6$) is 
given by one half the Euler characteristic, 
\beq
\chi={1\over{\vert{P}\vert}}\sum_{g,h\in P}\chi(g,h),
\label{euler}
\eeq
where $\chi(g,h)$ is the number of points left fixed 
simultaneously by $h$ and $g$. The mass formula for the right--movers in the
twisted sectors is given by, 
\beq
M_R^2=-1+{{(P+V)^2}\over2}+\Delta c_\theta+N_R
\label{mr2}
\eeq
where $V^I$ are the shifts on the gauge sector and 
$\Delta c_\theta={1\over4}\sum_{k}\eta_k(1-\eta_k)$ is the contribution of 
the twisted bosonic oscillators to the zero point energy and $\eta_k={1\over2}$
for a $Z_2$ twist.

To translate the fermionic boundary conditions to twists and shifts in the 
bosonic formulation we bosonize the real fermionic degrees of freedom, 
$\{y,\omega\vert{\bar y},{\bar\omega}\}$. Defining, 
${\xi_i}={\sqrt{1\over2}}(y_i+i\omega_i)=-ie^{iX_i}$, 
$\eta_i={\sqrt{1\over2}}(y_i-i\omega_i)=-ie^{-iX_i}$
with similar definitions for the right movers $\{{\bar y},{\bar\omega}\}$
and $X^I(z,{\bar z})=X^I_L(z)+X^I_R({\bar z})$. 
With these definitions the world--sheet supercurrents in the bosonic 
and fermionic formulations are equivalent,
\beq
T_F^{int}=\sum_{i}\chi_i{y_i}\omega_i=i\sum_{i}\chi_i{\xi_i}\eta_i=
\sum_{i}\chi_i\partial{X_i}.
\label{supercurrent}
\eeq
The momenta $P^I$ of the compactified scalars
in the bosonic formulation are identical with the $U(1)$ charges 
$Q(f)$ of the unbroken Cartan generators of the four dimensional
gauge group, 
\beq
Q(f)={1\over2}\alpha(f)+F(f)
\label{qf}
\eeq
where $\alpha(f)$ are the boundary conditions of complex fermions $f$,
reduced to the interval $(-1,1]$ and $F(f)$ is a fermion number operator. 

The boundary condition vectors $b_1$ and $b_2$ now translate into 
$Z_2\times Z_2$ twists on the bosons $X_i$ and fermions $\chi_i$ and 
to shifts on the gauge degrees of freedom. The massless spectrum
of the resulting orbifold model consist of the untwisted sector and 
three twisted sectors, $\theta$, $\theta^\prime$ and $\theta\theta^\prime$. 
From the untwisted sector we obtain the generators of the 
$SO(4)^3\times E_6\times U(1)^2\times E_8$ gauge groups. 
The only roots of $SO(12)$ that are invariant under the $Z_2\times Z_2$
twist are those of the subgroup $SO(4)^3$. Thus, the $SO(12)$ symmetry
is broken to $SO(4)^3$. Similarly, the shift in the gauge sector 
breaks one $E_8$ symmetry to $E_6\times U(1)^2$. In addition to the gauge group
generators the untwisted sector produces: three copies of $27+{\bar{27}}$,
one pair for each of the complexified NSR left--moving fermions; 
three copies of, $1+{\bar1}$, $E_6$ singlets which are charged under 
$U(1)^2$. These singlets are the untwisted moduli of the 
$Z_2\times Z_2$ orbifold model and match the number of 
untwisted moduli in the free fermionic model. 
The $E_8\times E_8$ singlets are obtained from the root lattice 
of $SO(12)$ and transform as $(1,4,4)$ under the $S0(4)^3$ symmetries, 
one for each of the complexified NSR left--moving fermions. 

The number of fixed points in each twist is 32. The total number of 
fixed points is 48. The number of chiral 27's is 24, eight from each
twisted sector, and matches the number of chiral 27's in the fermionic
model. For every fixed point we obtain the $SO(4)^3\times E_6\times E_8$
singlets. These are obtained for appropriate choices of the momentum 
vectors, $P^I$, and correspond to twisted moduli. 
The $E_6\times E_8$ singlets can be obtained by acting on the 
vacuum with twisted oscillators and from combinations of the dual of the 
invariant lattice, $I^*$, \cite{NSV}. 
The spectrum of the orbifold model and its symmetries are 
seen to coincide with the spectrum and symmetries of the
fermionic model \cite{foc}. 

It is noted that the effect of the additional basis vector $X$, Eq.
(\ref{vectorx}), is to separate the gauge degrees of freedom, spanned by
the world--sheet fermions $\{{\bar\psi}^{1,\cdots,5},
{\bar\eta}^{1},{\bar\eta}^{2},{\bar\eta}^{3},{\bar\phi}^{1,\cdots,8}\}$
from the internal compactified degrees of freedom $\{y,\omega\vert
{\bar y},{\bar\omega}\}^{1,\cdots,6}$. in the realistic free fermionic
models this is achieved by the vector $2\gamma$ \cite{foc}, with
\beq
2\gamma=(0,\cdots,0\vert{\underbrace{1,\cdots,1}_{{\bar\psi^{1,\cdots,5}},
{\bar\eta^{1,2,3}} {\bar\phi}^{1,\cdots,4}} },0,\cdots,0)~,
\label{vector2gamma}
\eeq
which breaks the $E_8\times E_8$ symmetry to $SO(16)\times SO(16)$.

\setcounter{footnote}{0}
\section{Discussion}

In section (\ref{higgs}) it was shown that the assignment of
boundary conditions to the set of world--sheet fermions $\{y,\omega\vert
{\bar y},{\bar\omega}\}^{1,\cdots,6}$ is the one that selects between
the electroweak Higgs doublet, versus the color Higgs triplets, according
to the quantity $\Delta_j$ in Eq. (\ref{dts}). In section (\ref{orbifold})
on the other hand it was shown that this set of world--sheet fermions
corresponds to the internal compactified dimensions in an orbifold
construction. This means that the assignment of boundary conditions
to this set of world--sheet fermions in fact corresponds to further
$Z_2$ orbifold twisting of the compactified dimensions. This type of
construction is precisely the one that has been recently rediscovered
in \cite{toons}. It is therefore very rewarding to note that in the
context of the heterotic--string construction such a mechanism operates
in a framework of realistic three generation models, compatible with
perturbative quantum gravity. It ought to be further remarked that
elimination of color triplets by Wilson line breaking has also
been noted in other orbifold compactifications \cite{ibanez}.

\bigskip
\medskip
\leftline{\large\bf Acknowledgements}
\medskip

I would like to thank the CERN theory group for hospitality.
This work was supported in part by PPARC.


\vfill\eject

\bibliographystyle{unsrt}

\end{document}